\documentclass[showpacs,preprint,aps]{revtex4}
\begin{document}
%
\title{
Self-assembled guanine ribbons as wide-bandgap semiconductors
}
\author{   
A.\ Calzolari, R.\ Di Felice,
E.\ Molinari, }
\affiliation{
Istituto Nazionale per la Fisica della Materia (INFM),
Universit\'a di Modena e Reggio Emilia, Via Campi 213/A, 41100 Modena, Italy
}
\author{
A.\ Garbesi}
\affiliation{
ISOF, Area della Ricerca CNR, Via P. Gobetti 101, 40129 Bologna, Italy
}
\begin{abstract}
We present a first principle study about the stability and the electronic properties of a new
biomolecular solid-state material, obtained by the self-assembling of guanine (G) molecules.
We consider hydrogen-bonded planar ribbons in isolated and stacked configurations. These aggregates 
present electronic properties similar to inorganic wide-bandgap semiconductors. The formation of
Bloch-type orbitals is observed along the stacking direction, while it is negligible in the ribbon plane.
Global band-like conduction may be affected by a dipole-field which spontaneously arises along the
ribbon axis. Our results indicate that G-ribbon assemblies are promising materials for biomolecular
nanodevices, consistently with recent experimental results.
\end{abstract}
\maketitle
%

The possibility that biomolecules may conduct current is  intriguing in the development
of molecular electronics. In particular, because of their sequence-specific properties, 
DNA molecules and their derivatives
are attractive for nanometer scale electronics, where their self-assembling capability might be used to wire
electronic-materials in a programmable way \cite{braun}, as well as to form novel conducting materials
 \cite{danny,rinaldi,rinaldi1}.\par
Indeed, in recent experiments \cite{rinaldi,rinaldi1}, a semiconducting behavior was displayed by a 
device constituted of a metallic
nanogate filled with a dried solution of a  lipophilic deoxyguanosine 
derivative.
This material is composed of hydrogen-bonded planar ribbons, which tend to stack and form solid-state fibers.
On a scale of $\sim100$ nm, the ribbons self-organize in ordered structures, giving rise to a film of nanocrystalline  
character \cite{rinaldi,rinaldi1,gottarelli}.
The results of transport and photocurrent experiments identify different regimes, 
depending on the width ($L$) of the nanogate:
when $L\sim100-200$ nm, few nanoscrystals are present between the electrodes and the I-V characteristics present the
typical metal-semiconductor-metal behavior; on the contrary, when the gap is reduced 
($L<100$ nm), a single crystal is trapped between electrodes, and the device acts as a rectifying diode.
Two concurrent mechanisms have been proposed to explain electron transport in such devices:
band-like conduction within the single crystals, and incoherent hopping among neighboring grains.
Band conduction and hopping mechanism have been also invoked to describe charge migration in other guanine-rich systems ({\sl e.g.}
G-quartets) or  poly(dG)-poly(dC) DNA \cite{danny,barton,schuster,giese,jortner}; 
however, no conclusive agreement was reached so far in this respect.
\par
In this paper, we present a first principle study of the electronic and conduction properties of self-assembled
guanine ribbons: using an extended-solid approach, we analyze the effects of H-bonding 
and $\pi$-$\pi$ coupling, that control
the microscopic interactions along the ribbon axis and along the stacking direction, respectively.
\noindent
Our method is based on {\sl ab-initio} molecular dynamics simulations: we performed total energy and 
force calculations, optimizing at the same time the atomic positions and the single-particle electronic
wave-functions, according to the Car-Parrinello approach \cite{fhi}.
The electronic structure is described within the Density Functional Theory (DFT) \cite{dft};
Generalized Gradient Approximation (BLYP-GGA) \cite{blyp} is applied in exchange-correlation functional to
allow an accurate description of H-bonding. The atomic potentials are described
by {\sl ab-initio} soft pseudopotentials in the Troullier-Martins \cite{mt} formulation.
The electronic wave functions are expanded in a plane-wave basis with an energy cutoff of 50 Ry.
\par
With respect to real deoxyguanosine ribbons,  our model leaves out the sugar (deoxyribose) and 
the attached paraffinic chains, because they are not involved in transport phenomena \cite{barton1,ladik}.
Since guanine  costitutes the building block of the ribbon aggregates, 
we first study the geometry and the electronic structure of the isolated molecule (see Figure 1, inset). 
Our method provides an accurate description of the structural parameters: 
bond lengths and angles are
reproduced with an error smaller than 2\% with respect to X-ray experimental data \cite{saenger}. 
The total charge density has a highly asymmetric distribution,
which induces a net dipole ($|\bar{\mu}|=7$ D) oriented as shown in the inset of Figure 1. 
The analysis of the electronic structure reveals
$\sigma$ and $\pi$ electron states.
The out-of-plane nature of $\pi$ orbitals, in particular of the  HOMO (Fig. 2a) and of the LUMO, may control 
the interactions in the stacking direction.
\par
To simulate extended ribbons, we assembled coplanar G molecules according to the experimental geometry \cite{gottarelli}.
By optimizing the structure of an isolated G dimer, we obtained the intermolecular distance. We used this value to fix the periodicity of the extended structures, whose geometry was also optimized.
The periodicity of the ribbons (1.1nm) is in good agreement with the experimental data (1.2nm).
\noindent
The model ribbons (Fig. 1) are 1D planar wires of guanines connected by a H-bond network. 
The donor-acceptor distances (NH$\cdots$ N=2.9 \AA, NH$\cdots$ O=2.8 \AA) are consistent
with previous theoretical results for similar systems ({\sl e.g.} DNA base-pairs) \cite{sponer}. 

\noindent
H-bonding is a  weak interaction, not capable of modifing the internal configuration of the single guanine;
we find that its main effect is to stabilize the structure, as proved by  large calculated adsorption energy
($\Delta E=-810$ meV/G).
On the other hand,  H-bonding does not favor the lateral coupling of electron states between neighboring
molecules. 
As shown in  Figure 2b, where the charge density plot of the HOMO is presented,
the ribbon states are localized around the single guanines, with no Bloch-type delocalization through
the hydrogen-bonds.
The density of states (DOS, top of Fig. 3a) shows isolated sharp peaks corresponding to the energy levels of G, and
the energy bands (not shown here) are dispersionless. We thus conclude that band-like conductivity, through
extended states, is not a viable mechanism
for electron transport along the ribbon axis.
\par
The picture changes drastically when we consider stacked configurations. 
In a previous work \cite{prb}, we demonstrated that
the electronic properties of vertical stacks of G molecules are very sensitive to the $\pi - \pi$ coupling:
A strong  $\pi - \pi$ superposition is mandatory to have a Bloch-like conduction
along the stacking direction. Here, we present a stacked configuration of planar ribbons, where the stacking
direction is orthogonal to the plane of the ribbon. 
Even though experimental results prove that the ribbons self-organize in fibers 
 \cite{rinaldi}, the  X-ray structure of single crystal is still missing, therefore
the real stacking geometry is unknown. We report our results for a  configuration in which the ribbons are perfectly
eclipsed one above each other and the band dispersion is expected to be maximized.
In this configuration, we  investigated the concurrent effects of H-bonding in the 
ribbon plane and  $\pi - \pi$ coupling along the stacking direction: a reduced value of the adsorption energy 
($\Delta E=-310$ meV/G), with respect the isolated wire, is the index of 
electrostatic repulsion of $\pi - \pi$ superposition.\\
We then examined the electron states near the gap edge, that affect the conduction properties;  
these are $\pi$ orbitals very sensitive to the vertical superposition. As an example,
the LUMO of the eclipsed structure is shown in Fig. 2c: 
the $\pi - \pi$ coupling induces the formation of Bloch-type orbitals delocalized  along stacking direction.
On the other hand, as it happens for isolated ribbons, the electronic states are not extended in the plane, 
indicating that
H-bonding gives negligible contribution to band conductivity.
This  is consistent with the bandstructure (Fig. 3b).
The HOMO- and LUMO-derived bands are dispersive along the stacking direction ($\Gamma - A$); 
while they are almost flat along the ribbon axis ($\Gamma - X$).
The calculated effective masses for electrons and holes are 
$m_{e}=1.24$ and $m_{h}=1.07$ (in units of the free electron mass)
along the stacking direction, whereas they are 
very high along the axis of the ribbons.
We conclude that the
$\pi - \pi$ interaction enhances band properties, improving the mobility of the charge carriers
only along the stacking direction.
The DOS (bottom of Fig. 3a) for the eclipsed ribbons confirms this behavior, 
and shows wider ranges of allowed energies with respect to isolated ribbons, expecially in the region around the gap. 
Band properties of fibers may be similar to those of inorganic wide-bandgap semiconductors ({\sl e.g.} GaN, AlN \cite{pugh})
for which Bloch conduction has been proved in presence of doping.
Our results are in agreement with the experimental evidence of a semiconducting behavior of guanine ribbons.
\par
In order to explain the rectifying properties of single nanocrystals, 
we must consider the polar nature of the G-ribbons.
Due to  the relative orientation of the single G's in the ribbons, molecular dipoles arrange  to form a net dipole
oriented along the ribbon axis (Fig. 1). 
The calculated values of  dipoles, for finite sequences of 2 ($|\bar{\mu}|=14.6$ D) 
and 4 ($|\bar{\mu}|=32.8$ D) guanines, indicate that a dipole-dipole interaction arises along ribbons.
The associated electric field may couple with proper Bloch-momentum of charge carriers, defining a different effective 
direction for electron transport. The presence of a favorite direction justify the diode-like character of experimental
I-V curves.
\par 
In summary, we have demonstrated that supramolecular aggregates of guanines may form solid-state materials with interesting
conduction properties.
Ribbon fibers present wide-bandgap semiconducting behavior, compatible with Bloch-like conduction along stacking
direction. A strong dipole field influences band properties of such nanocrystals, 
inducing preferential direction in movement of electron and holes, as confirmed by recent experimental results.
 \cite{rinaldi}.

%

{\bf Acknowledgements}\\
Funding was provided by INFM through 
{\sl Progetto Calcolo Parallelo}, and PRA-SINPROT.

%
\begin{figure}
\caption{Isolated planar guanine ribbon:   Relaxed geometry, dipole moment and isosurface of total charge density.
In the inset the relaxed geometry, the dipole moment and the isosurface of total charge density of a single guanine molecule are shown.}
\end{figure}
\begin{figure}
\caption{a) Isosurface of the HOMO  for the isolated G molecule; 
b) Isosurface of the HOMO for the planar guanine ribbon;
c) Isosurface of the LUMO for stacked planar ribbons in the eclipsed configurations.}
\end{figure} 
\begin{figure}
\caption{a)Density of States (DOS) for isolated (top) and stacked (bottom) planar ribbons;
b)Bandstructure for stacked planar ribbons in the eclipsed configuration.}
\end{figure}
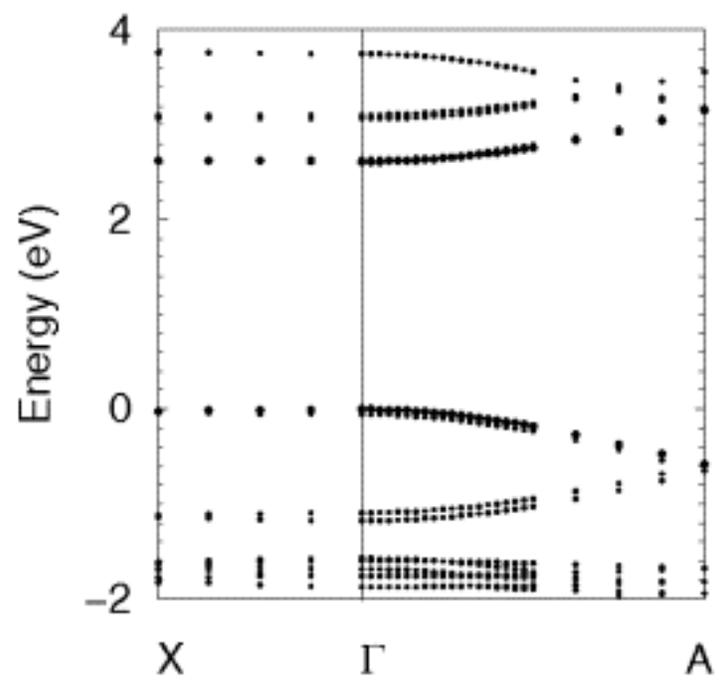
\end{document}